# Comment on "A new nanoscale metastable iron phase in carbon steels"

Cyril Cayron

Laboratory of ThermoMechanical Metallurgy (LMTM), PX Group Chair, Ecole Polytechnique Fédérale de Lausanne (EPFL), Rue de la Maladière 71b, 2000 Neuchâtel, Switzerland.

cyril.cayron@epfl.ch

## Abstract

We show that the selected area diffraction patterns presented in a recent paper [1] do not prove the existence of a new hexagonal phase in martensitic steels. They can be actually simulated by twin effects.

Very recently a paper called "*A new nanoscale metastable iron phase in carbon steels*" was published in *Scientific Reports* by Liu et al. [1]. In this paper, the authors used Selective Area Diffraction (SAED) patterns in Transmission Electron Microscopy (TEM) to claim the existence of a hexagonal phase that they called $\omega$ and that they consider to be an intermediate phase for the (face-centered cubic) fcc to (body-centered cubic) bcc martensitic transformation in steels. We will show that all the SAED patterns presented in this paper can be actually explained by twinning artifacts (double diffraction and streaking). Since twins are well known in iron carbon steels, it seems legitimate to conclude that the $\omega$ phase does not exist in the steels studied by the authors.

It is not the first time, and it seems that it will not be the last time, that scientists are confused by the twin artifacts in electron diffraction and claim the existence of a "new" hexagonal phase. In 2009, we tried to warn of such confusion in silicon nanowires [2]. Literature shows a long story of alternate publications of papers, (a) some claiming the existence of new hexagonal phases, and (b) others infirming the conclusions and warning of the effect of twin artifacts (see references in [2]). It seems that the papers of type (b) are regularly forgotten, which lets papers of type (a) regularly flourish at the same rhythm. The paper of Liu et al [1] is of type (a), and our comment will be of type (b).

In order to simulate the twin effects, we have used a computer program called GenOVa [3] that allow us to generate the twin variants, as we did in our work on silicon nanowires [2]. This software includes the simulation of double diffraction and the streaking effects. The positions of the streaks in the SAED patterns are determined by calculating the intersection of the streaks originated from the diffraction spots of high order Laue zones with the Ewald sphere. In the present work, we have simulated the SAED patterns of paper [1] by a using a classical bcc structure, with (i) the double diffraction between the $\Sigma 3$ macrotwins, and (ii) the streak effects along the normal to the $\{211\}_{bcc}$ planes which are the habit planes of the nanotwins. The figures of the present paper are numbered with Romain numerals in order to distinguish them from the figures of work [1].

Let us start by the SAED pattern of Fig.1. The authors could simulate it by twins, but that solution was rejected because the "*the incident electron beam is parallel with the twinning boundary plane and there is no overlap between the matrix and the twin in the depth direction, it is normally impossible to observe any double diffraction in the twinning diffraction pattern*". It is true that the $\{\bar{2}1\bar{1}\}$ twin plane is oriented along the electron beam which is along the [011] zone axis, and it is probable that the twinned bcc crystals do not overlap, but a dynamical effect is actually possible laterally due to the very small dimensions of the twins. Indeed, double diffraction comes from the deviation of the incident electron beam diffracted by the first crystal, which is then diffracted again by the second twinned crystal. If the second crystal is far from the first crystal, the diffracted beam can't reach the second crystal, and a double diffraction is impossible; but for the intricate nanotwins it is possible. Indeed, the diffraction angles are of order of 10-20 mrad, and the thickness of the TEM samples are around 100 nm, which means that the diffracted beam is deviated lateraly by 1-2 nm, i.e. the order of magnitude of the size of the nanotwins shown in the HRTEM of Fig1d. If one also considers that the electron beam is not exactly orientated along the zone axis and the twin plane is slightly tilted, then the hypothesis of double diffraction becomes very plausible. The simulation is given in Fig. I.

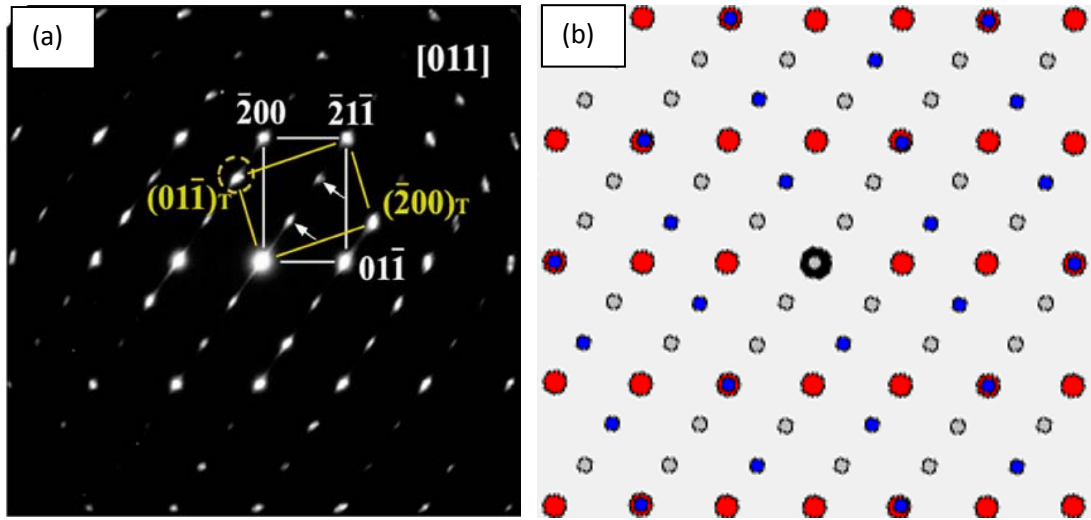

*Fig. I.  Simulation of the SAED of Fig.1 of paper [1]. (a) Experimental SAED, (b) simulation of (a) with the electron beam along the [011] zone axis, with the spots of the bcc crystal represented by the red disks, the spots of the twinned bcc crystal by the blue disks, and the double diffraction by the grey disks.*

The authors have also rejected the possibility that the extra spots in the SAED pattern of Fig4a could come from twins by arguing that : "*These extra spots cannot be treated as double diffraction of the {112}<111>-type twinning structure because the twinning plane is {112}, which is perpendicular to the incident electron beam. In such cases, all matrices and twins will result in exactly the same diffraction spots and the diffraction patterns should be totally overlapped.*" That is true, but the authors did not consider the streaking effects perpendicularly to the other {112} nanotwin planes, i.e. those which do not contain the zone axis. This effect actually fully explains the extra spots of Fig4a, as shown in Fig. II.

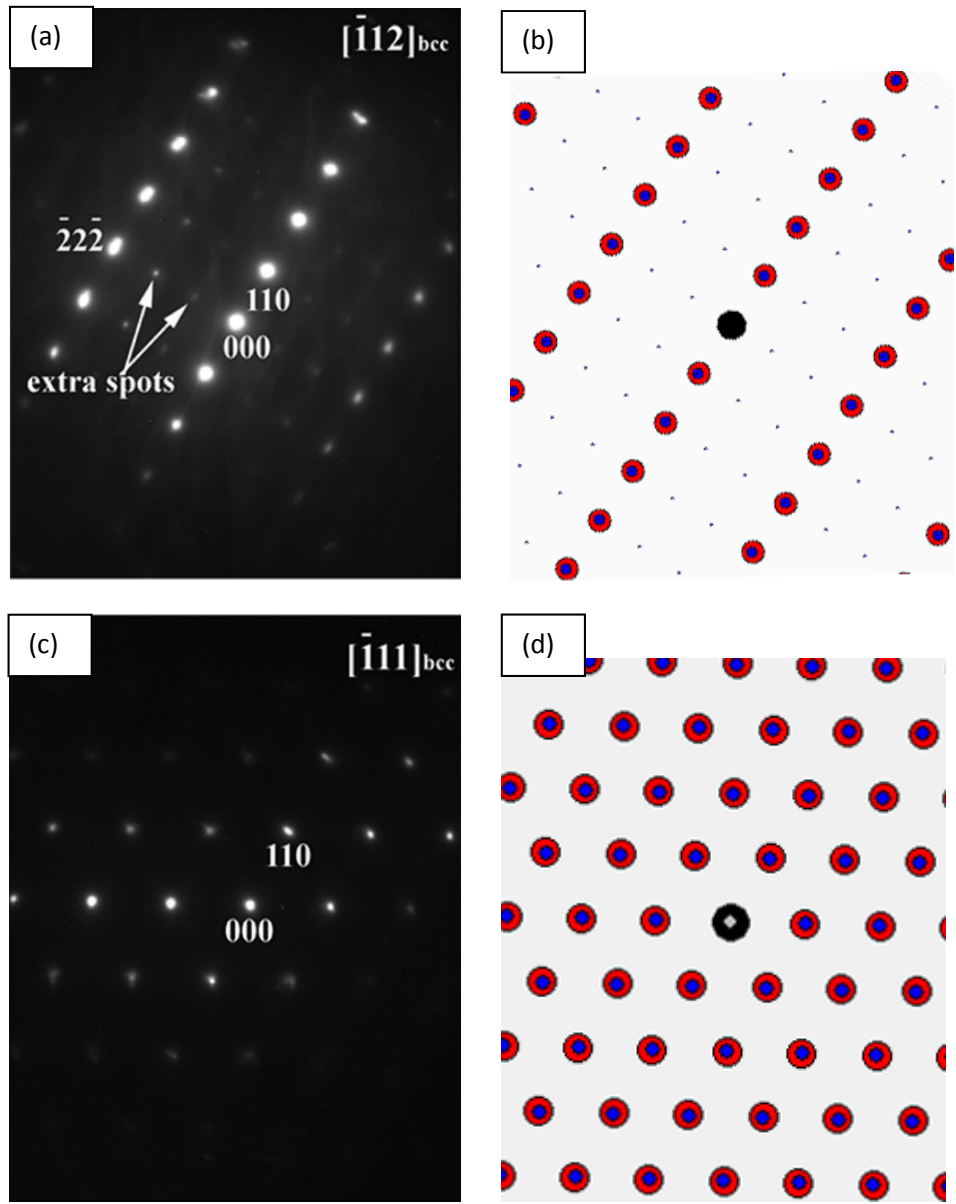

*Fig. II.    Simulation of the SAED of Fig.4 of paper [1]. (a-c) Experimental SAED. (b) Simulation of (a) with the zone axis along [1̄12]; the spots of the bcc crystal are represented by the red disks, the spots of the twinned bcc crystal by the blue disks, and the extra spots coming from a streaking effect of the higher Laue zone along the [112] direction by the black spots. (d) Simulation of (c) with the same configuration of twins, but now oriented along the [1̄11] zone axis. The extra spots are no more visible because they intercept the Ewald sphere at the same positions as the existing spots of the bcc crystal.*

The last SAED pattern that remains to simulate is the one of Fig.7. As all the others, it can be explained by twin effects, as shown in Fig. III.

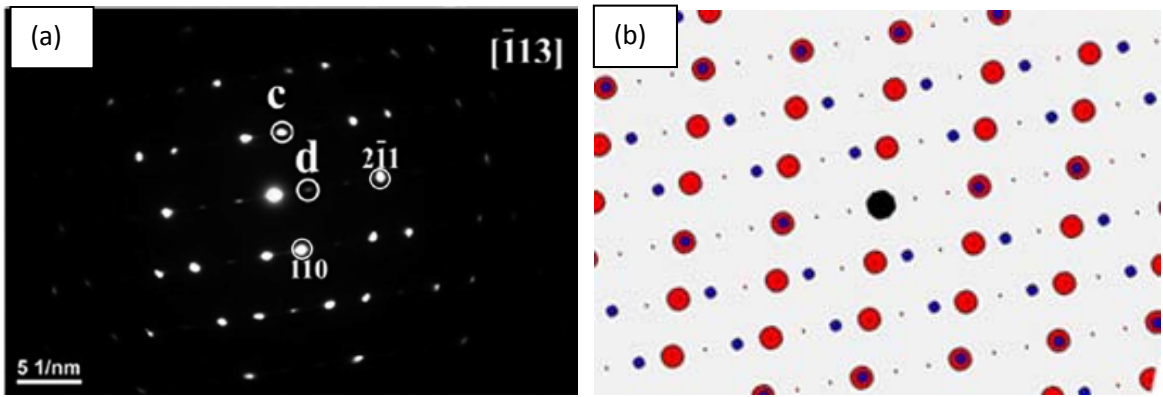

*Fig. III.  Simulation of the SAED of Fig.7. (a) Experimental SAED, (b) simulation of two twinned crystal oriented along [1̄13] (the red disks) and [113] (the blue disks. The extra spots originated by streaks of the higher Laue zones along the [211] direction are given by the black dots.*

We note that the SAED patterns of Fig. 8 were simulated by twin effects by the authors themselves (Fig.8 d,e,f).

All the SAED patterns presented in paper [1] can be simulated with a bcc structure and $\Sigma 3$ {112} macro and nanotwins taking into account double diffraction and streaking perpendicularly to the {112} planes. Therefore, it should be concluded that the SAED patterns do not prove the existence of a new hexagonal phase. We should have been pleased if such a hexagonal phase could have been put in evidence because it would have confirmed an earlier "two-step" model of fcc-bcc martensitic transformation that we have imagined few years ago [4], even if we have modified this model to develop a continuous version [5]. We share the same aim of the authors of paper [1] and we also research a "universal" link between the phase transformations in different metallic system, but we do not think anymore that this link should be given by a hypothetical intermediate hexagonal phase. We believe that the unifying link between the fcc, hcp and bcc phases is the nature of the lattice distortion [6].

We agree with some observations pointed by the authors when they note that mechanical twins should be straight whereas in their images, as in many other TEM images reported in literature, the twins appear curved. We agree with them to say that these twins are not mechanical twins. However, to our point of view, they do not either correspond to the "new" ω phase. Actually, there are 24 KS variants of martensite, and the variants are not distributed randomly. They are grouped by pairs or more complex sets [7]-[9]. There is especially the case for $\Sigma 3$ twin pairs. So, it is highly probable that the twinned bcc crystals are pairs of two Kurdjumov-Sachs martensite variants in twin relationship.

In conclusion, the simulations have proved that the SAED patterns shown in paper [1] can be explained by twinning effects (double diffraction and streaking). Therefore, there is no experimental proof of the existence of the ω phase in paper [1].